\documentstyle[10lomcon,cite,psfig]{article}
\newcommand{\be}{\begin{equation}}
\newcommand{\ee}{\end{equation}}
\newcommand{\ds}{\displaystyle}

\newcommand{\vpint}{\int\makebox[0mm][r]{\bf --\hspace*{0.13cm}}}
\newcommand{\vph}{\varphi}

\begin{document}

\title{QCD$_2$ IN THE AXIAL GAUGE REVISITED}

\author{ Yu.S.Kalashnikova \footnote{e-mail: yulia@heron.itep.ru}}
\address{ITEP, 117218, B.Cheremushkinskaya 25, Moscow, Russia}
\author{A.V.Nefediev \footnote{e-mail: nefediev@heron.itep.ru}}
\address{ITEP, 117218, B.Cheremushkinskaya 25, Moscow, Russia\\and\\
Centro de F\'\i sica das Interac\c c\~oes Fundamentais (CFIF)\\
Departamento de F\'\i sica, Instituto Superior T\'ecnico\\
Av. Rovisco Pais, P-1049-001 Lisboa, Portugal}

\maketitle\abstracts{
The 't Hooft model for the two-dimensional QCD in the limit of infinite
number of colours is studied in the axial gauge. The mass-gap and
the bound-state equations are derived using the two consequent
Bogoliubov-like transformations. Chiral properties of the model are
studied in the Hamiltonian and matrix approaches to the latter. Special
attention is payed to the explicit pionic solution of the bound-state
equation.}

The model for the two-dimensional QCD in the limit of infinite number of colour was
suggested in 1974 \cite{1} and it is a marvelous playground for testing methods and
approaches used in QCD$_4$. It is described by the Lagrangian
\be
L (x)= -\frac{1}{4}F^a_{\mu\nu}(x)F^a_{\mu\nu}(x) + \bar q(x)(i\gamma_{\mu}
\partial_{\mu} - gA^a_{\mu}t^a\gamma_{\mu} - m)q(x),
\label{LL}
\ee
whereas the large $N_C$ limit implies that $\gamma\equiv\frac{g^2N_C}{4\pi}$ remains
finite, so that a nontrivial set of diagrams (planar diagrams) appear to be of the
same order in $N_C$ and should be summed. 

Following \cite{2,4} we consider the model (\ref{LL}) in the axial gauge $A_1^a(x_0,x)=0$ and
use the principal-value prescription to regularize the infrared divergences. We also
define the dressed quark field
\be
q_{\alpha i}(t,x)=\int\frac{dk}{2\pi}e^{ikx}[b_{\alpha}(k,t)u_i(k)+d_{\alpha}^+
(-k,t)v_i(-k)],
\ee
\be
\ds\{b_{\alpha}(t,p)b^+_{\beta}(t,q)\}=
\ds\{d_{\alpha}(t,-p)d^+_{\beta}(t,-q)\}=2\pi\delta(p-q)\delta_{\alpha\beta},
\ee
\be
u(k)=T(k)\left(1 \atop 0 \right),\quad v(-k)=T(k)\left(0 \atop 1 \right),\quad
T(k)=\exp{\left[-\frac{1}{2}\theta(k)\gamma_1\right]},
\ee
where the Bogoliubov angle $\theta(p)$ is the solution to the mass-gap equation.
To derive the latter, we organize the normal ordering of the Hamiltonian in the new
basis,
\be
H= LN_C{\cal E}_v + :H_2: + :H_4:,
\label{H1}
\ee
and demand that the quadratic part be diagonal in terms of the dressed-quark creation
and annihilation operators. This implies that $\theta$ is a solution to the following
equation:
\be
p\cos\theta(p)-m\sin\theta(p)=\frac{\gamma}{2}\vpint\frac{dk}{(p-k)^2}\sin[\theta(p)-\theta(k)],
\label{mge}
\ee
and the dressed-quark dispersive law is
\be
E(p)=m\cos\theta(p)+p\sin\theta(p)+\frac{\gamma}{2}\vpint\frac{dk}{(p-k)^2}\cos[\theta(p)-\theta(k)].
\ee

An alternative way to derive the mass-gap equation (\ref{mge}) is to minimize
the vacuum energy (the first term on the r.h.s. in (\ref{H1})),
\be
{\cal E}_v =\int\frac{dp}{2\pi}Tr
\left\{\gamma_5p\Lambda_{-}(p)+\frac{\gamma}{4\pi}
\int\frac{dk}{(p-k)^2}\Lambda_{+}(k)\Lambda_{-}(p)\right\},
\label{Evac}
\ee
with the projectors being $\Lambda_{\pm}(p)=T(p)\frac{1\pm\gamma_0}{2}T^+(p)$.

In the meantime, one should be extremely careful with the divergences which show up
in the second term of (\ref{Evac}). Indeed, defining the difference between the
vacuum energy of a nontrivial vacuum and the free-theory one,
\be
\Delta{\cal E}_v[\theta]={\cal E}_v[\theta]-{\cal E}_v[\theta_{\rm free}],\quad
\theta_{\rm free}(p)=\frac{\pi}{2}{\rm sign}(p),
\label{Evac2}
\ee
one can use the following trick \cite{BRN}. If $\theta(p)$ is the function which
provides the minimum of the vacuum energy (\ref{Evac2}), then the function with the
same profile, but stretched with an arbitrary parameter $A$, 
$\theta(p)\to\theta(p/A)$,
should enlarge the energy. A simple analysis demonstrates, that this is, indeed, the
case, since $\Delta{\cal E}_v$ has the following behaviour as a function of $A$:
\be
\Delta{\cal E}_v=\frac12C_1A^2-\gamma C_2\ln A+\gamma C_3
\label{Evac3}
\ee
with $C_{1,2,3}$ being positive constants. The logarithm in the second term on the
r.h.s. of (\ref{Evac3}) is due to the infrared divergence of the integral in
(\ref{Evac}), and the constant $C_3$ contains the logarithm of the cut-off. The
function (\ref{Evac3}) has a minimum for a nonzero $A$ and, what is more, there is no
minimum for $A=0$ provided $\gamma\neq 0$. The latter fact means that there is no
phase of the theory with the angle $\theta=\frac{\pi}{2}{\rm sign}(p)$ found in
\cite{2}, which
leads to zero chiral condensate and unbroken chiral symmetry. Thus the 't~Hooft model has
the only nontrivial vacuum which provides spontaneous breaking of the chiral symmetry
and which is defined by the numerical solution of equation (\ref{mge}) found in
\cite{5}. Formula (\ref{Evac3}) can be written in a more physically transparent form
if one notices that the chiral condensate transforms linearly under the
above-mentioned transformation $(\Sigma\equiv \langle \bar q q\rangle\to A\Sigma)$,
so that one can use $\Sigma$ instead of $A$:
\be
\Delta{\cal E}_v=C_1'\left[\frac12\left(\Sigma/\Sigma_0\right)^2-
\ln\left|\Sigma/\Sigma_0\right| \right]+\gamma C_3'
\ee
with $\Sigma_0$ being the real condensate of the model which will be discussed below.

From the original paper \cite{1} it is known that the "physical" degrees of
freedom of the model are the quark-antiquark mesons. The operators creating and
annihilating mesonic states can be defined via the dressed quark states as \cite{KNV1}
\be
m_n=\int\frac{dq}{2\pi\sqrt{N_C}}\left\{d_i(P-q)b_i(q)\vph_+^n(q,P)+
b^{+}_i(q)d^{+}_i(P-q)\vph_-^n(q,P)\right\},
\label{m}
\ee
where the two wave function appeared for each meson. The "+" function describes
the motion of the $q\bar q$ pair in meson forward in time, whereas the "-" one
describes the backward motion. The orthogonality and completeness conditions for
$\vph$'s contain the negative sign between the "+" and the "-" parts, $e.g.$,
\be
\begin{array}{rcl}
\ds\int\frac{\ds dp}{\ds 2\pi}\left(\vph_+^n(p,Q)\vph_+^{m}(p,Q)-\vph_-^n(p,Q)\vph_-^m(p,Q)
\right)&=&\delta_{nm},\\
&&\\
\ds\int\frac{dp}{2\pi}\left(\vph_+^n(p,Q)\vph_-^{m}(p,Q)-\vph_-^n(p,Q)\vph_+^m(p,Q)
\right)&=&0.
\label{norms}
\end{array}
\ee
This sign is a consequence of the second Bogoliubov-like transformation which is to
be performed in the theory to find the form (\ref{m}) of the mesonic operators, 
so that the two components of the mesonic wave function play the role of
the standard Bogoliubov amplitudes $u$ and $v$ \cite{KNV1}. The Hamiltonian of the model takes
the diagonal form in the new basis,
\be
H=LN_C{\cal E}'_v
+\sum\limits_{n=0}^{+\infty}\int\frac{dP}{2\pi}P^0_n(P)m^+_n(P)m_n(P)+
O\left(\frac{1}{\sqrt{N_C}}\right),
\ee
if the wave functions obey the bound-state equation in the form of a system of 
two coupled equations \cite{2,4}:
\be
\left\{
\begin{array}{c}
[K(p,P)-P_0]\vph_+(p,P)=\gamma\ds\vpint\frac{\ds dk}{\ds (p-k)^2}
\left[C\vph_+(k,P)-S\vph_-(k,P)\right]\\

[K(p,P)+P_0]\vph_-(p,P)=\gamma\ds\vpint\frac{\ds dk}{\ds (p-k)^2}
\left[C\vph_-(k,P)-S\vph_+(k,P)\right],
\end{array}
\right.
\label{BG}
\ee
with $K(p,P)=E(p)+E(P-p)$, $C=\cos\frac{\theta(p)-\theta(k)}{2}
\cos\frac{\theta(P-p)-\theta(P-k)}{2}$ and $S=\sin\frac{\theta(p)-\theta(k)}{2}
\sin\frac{\theta(P-p)-\theta(P-k)}{2}$.
Note that the system (\ref{BG}) can be written in the form of a Dirac-type equation
in the Hamiltonian form,
\be
\hat {\cal H}\psi=Q_0\psi,\quad
\hat {\cal H}=\left(
\begin{array}{cc}
K-\hat{C}&\hat{S}\\
-\hat{S}&-K+\hat{C}
\end{array}
\right)
=\gamma_0(K-\hat{C})+\gamma_1\hat{S},
\label{Hmat2}
\ee
where for an arbitrary function $F(p,P)$ operators $\hat{S}$ and $\hat{C}$ act as
\be
\hat{C}(\hat{S})F(p,P)\equiv \gamma\int\frac{dk}{(p-k)^2}C(S)(p,k,P)F(k,P).
\ee

The operator $\hat {\cal H}$ is not Hermitian, and the distorted form of the
norm (\ref{norms}) is a remnant of this fact. Nevertheless 
the positiveness of its eigenvalues can be proved explicitly \cite{KN2}. 

Using the Hamiltonian approach to the model developed above, one can study its chiral
properties, among which we mention
\begin{itemize}
\item The chiral pion --- the exact massless (in the chiral limit) solution of 
the bound-state equation (\ref{BG}) \cite{KN2}:
\be
\vph^{\pi}_{\pm}(p,Q) =\sqrt{\frac{\pi}{2Q}}\left(\cos\frac{\theta(Q-p)-\theta(p)}{2}\pm
\sin\frac{\theta(Q-p)+\theta(p)}{2}\right).
\ee
\item The pion decay constant, which can be defined for the 
above-mentioned state ($|\Omega\rangle$ is the vacuum annihilated 
by mesonic operators (\ref{m})):
\be
\left.\left\langle\Omega\right.\right|\bar{q}(x)\gamma_{\mu}\gamma_5q(x)\left|\left.\pi(Q)\right.
\right\rangle=f_{\pi}Q_{\mu}\frac{e^{-iQx}}{\sqrt{2Q_0}},\quad
f_{\pi}=\sqrt{\frac{N_C}{\pi}}.
\ee
\item Partial conservation of the axial-vector current, which holds true in the
operator form,
\be
J_{\mu}^5(x)=if_{\pi}\partial_{\mu}\Psi_{\pi}(x),\quad f_n=f_{\pi}\delta_{n\pi}.
\ee
\item The chiral condensate, 
\be
\langle\bar{q}q\rangle (m=0)=-N_C\int_{-\infty}^{+\infty}\frac{dp}{2\pi}\cos\theta(p)=
-0.29N_C\sqrt{2\gamma}.
\ee
\item The Gell-Mann-Oakes-Renner relation,
\be
f_{\pi}^2M_{\pi}^2=-2m\langle\bar{q}q\rangle,
\ee
which can be checked explicitly.
\end{itemize}

Now we turn to the properties of currents in the 't~Hooft model. The conservation
laws can be derived for both, vector and axial-vector, currents in the chiral limit,
which read:
\be
V^M_{\mu}(P)=\langle \Omega|\bar q\gamma_{\mu} q|M,P\rangle,\quad P_0^MV^M_0-PV^M=0,
\ee
\be
A^M_{\mu}(P)=\langle \Omega|\bar q\gamma_{\mu}\gamma_5 q|M,P\rangle,\quad
P_0^MA^M_0-PA^M =0.
\ee

For the current-quark-antiquark vertices one has the vector and the
axial-vector Ward identities \cite{Ein,KN2,4}
\be
-iP_{\mu}v_{\mu}(p,P)=S^{-1}(p)-S^{-1}(p-P),
\label{Wv}
\ee
\be
-iP_{\mu}a_{\mu}(p,P) = S^{-1}(p)\gamma_5+\gamma_5S^{-1}(p-P),
\label{Wa}
\ee
and, as a result, the following conservation laws:
\be
(P_{\mu}-P'_{\mu})\langle M,P|v_{\mu}|M',P'\rangle=0,
\ee
\be
(P_{\mu}-P'_{\mu})\langle M,P|a_{\mu}|M',P'\rangle=0.
\ee

In derivation of the latter formulae the matrix approach to the model was used
\cite{Ein,KN2}, in which an effective diagrammatic technique is defined containing 
the matrix form of the wave function obeying the matrix bound-state equation,
the dressed quark propagator $S(p)$, the dressed meson-quark-antiquark vertex $\Gamma$
($\bar\Gamma$ for outgoing meson), the dressed quark-quark scattering amplitude, and,
finally, the effective coupling constant $-i\gamma/\sqrt{N_C}$, which is to be
prescribed to each meson-quark-antiquark vertex. For example, for the pion one can
find the following vertex function:
\be
\Gamma_{\pi}(p,P)=S^{-1}(p)(1+\gamma_5)-(1-\gamma_5)S^{-1}(p-P),
\label{Gp1}
\ee
which can be easily identified with the divergences of the vector and axial-vector
currents (\ref{Wv}) and (\ref{Wa}):
\be
\Gamma_{\pi}(p,P)=-iP_{\mu}v_{\mu}(p,P)-iP_{\mu}a_{\mu}(p,P).
\label{Gp2}
\ee

It is not surprise that the pion couples to both, the vector and the axial-vector
currents, since in two dimensions the two currents are dual to one another. 
The interested reader can find details of the approach in the review paper \cite{4}.

As a next application of the Hamiltonian and the matrix approaches discussed above, let
us study hadronic decays in this theory. If meson $A$ decays into mesons $B$ and $C$,
then the amplitude of such a process can be found in two ways: in the Hamiltonian
technique,
\be
M(A \to B+C)=\langle B(P_B)C(P_C)|\Delta H|A(P_A)\rangle,
\label{amH}
\ee
or using the matrix approach, directly from the corresponding decay diagrams:
\be
M(A \to B+C)=-\frac{i\gamma^3}{\sqrt{N_C}}\int \frac{d^2k}{(2\pi)^2} 
Sp[\Gamma_A(k+P_B,P_A)S(k-P_C)
\label{amM}
\ee
$$
\times\bar\Gamma_C(k,P_C)
S(k)\bar \Gamma_B(k+P_B,P_B)S(k+P_B)]+(B \leftrightarrow C).
$$

Both equations, (\ref{amH}) and (\ref{amM}), lead to one and the same 6-term form of
the amplitude (see \cite{KN2,4} for details), which possesses contributions of both
components of the wave function for each meson. If one of the final states is the
pion and the chiral limit is used, then the explicit form of the vertex (\ref{Gp1}) 
can be substituted into
(\ref{amM}) that immediately leads to the result that the amplitude vanishes,
\be
M(A\to\pi+C)\equiv 0,
\label{M0}
\ee
which also follows from the identification (\ref{Gp2}) and the vector and
axial-vector conservation laws.

The result (\ref{M0}) could be anticipated, since the pion
decay constant is dimensionless in the 't~Hooft model. As a result, the Adler
selfconsistency condition for the amplitudes with pions involved \cite{Adler} has to
hold true for any pion momentum.

In conclusion let us say, that the two-dimensional 't~Hooft model possesses many
features which are known to be inherent to QCD$_4$ and it can be used as a test
laboratory to see how all these beautiful properties may appear.

\section*{Acknowledgments}
Financial support of RFFI grants 00-02-17836 and 00-15-96786,
INTAS-RFFI grant IR-97-232 and INTAS CALL 2000-110 
is gratefully acknowledged. One of the authors (A.N.) is also supported via RFFI grant
01-02-06273.

\end{document}